\documentclass[aps,twocolumn,amssymb]{revtex4}

\usepackage{graphicx}
\usepackage{multirow}
\usepackage[breaklinks,colorlinks,urlcolor=blue,citecolor=blue]{hyperref}
\usepackage{mathrsfs}
\usepackage{amsmath}
\usepackage{soul}
\usepackage{color}
\usepackage[normalem]{ulem}

\begin{document}

\title{Two- and Multi-dimensional Curve Fitting using
Bayesian Inference}
\author{Andrew W. Steiner$^{1,2}$}
\affiliation{$^{1}$Department of Physics and Astronomy, University of
  Tennessee, Knoxville, TN 37996, USA}
\affiliation{$^{2}$Physics Division, Oak Ridge National Laboratory, Oak
  Ridge, TN 37831, USA}
\begin{abstract}
  Fitting models to data using Bayesian inference is quite common, but
  when each point in parameter space gives a curve, fitting the curve
  to a data set requires new nuisance parameters, which specify the
  metric embedding the one-dimensional curve into the
  higher-dimensional space occupied by the data. A generic formalism
  for curve fitting in the context of Bayesian inference is developed
  which shows how the aforementioned metric arises. The result is a
  natural generalization of previous works, and is compared to
  oft-used frequentist approaches and similar Bayesian techniques.
\end{abstract}
\maketitle

Curve fitting, as opposed to a standard nonlinear fit, becomes
relevant when a data set has a significant uncertainty in the
direction of the independent variable (sometimes referred to as
errors-in-variables models) or the set of model curves includes
relations which predicts multiple values for the same value of the
independent variable. In this case, the standard fitting approaches,
such as a ``chi-squared analysis'' become unusable. In the context of
Bayesian inference, this work shows that one can construct a general
formalism for curve fitting which is applicable to a wide variety of
data analysis problems and is not yet published elsewhere. In this
Letter, a Bayesian approach to curve fitting is described and compared
to almost all previous works which have given similar techniques which
are less general.

Bayesian inference often proceeds via the application of Bayes
theorem, $P({\cal M}|{\cal D}) = P({\cal D}|{\cal M}) P({\cal M})/
P({\cal D}) $, and the process of fitting models to data begins with
the construction the conditional probability of the data given the
model, $P({\cal D}|{\cal M})$ (this will be referred to this as the likelihood
function) and specifying the prior probability distribution, $P({\cal
  M})$ (the probability distribution for the data, $P({\cal D})$, is
often not needed). Since the RHS of Bayes' theorem involves a product,
there is an obvious potential for ambiguity: a multiplicative factor
can be removed from the conditional probability and placed in the
prior distribution without any modification to the final result.


It is assumed that the set of (possibly correlated) data points may 
be represented by a single normalizable probability density function
${\cal D}(x_1,x_2,\ldots,x_N)$ over $N$ quantities of interest. As an
example, in the case that one has $N$ uncorrelated one-dimensional
data points which have normally distributed errors, then ${\cal D}$ is
a product of $N$ Gaussian distributions. In the case that these
one-dimensional data points have correlations in the form of a
multi-dimensional Gaussian (the formalism below does not require
this assumption), then
\begin{equation}
  {\cal D}(x_1,x_2,\ldots,x_N) \propto \mathrm{exp}
  \left[ - \frac{1}{2} (x_i-\mu_i) \Sigma_{ij}^{-1} (x_j-\mu_j) \right]
  \label{eq:mgauss}
\end{equation}
where $\vec{\mu}$ is the peak and $\Sigma$ is a covariance matrix.
In some cases, the quantities of interest can be grouped together into
n-tuples, so the quantities of interest are relabeled and the
data probability density distribution is rewritten
\begin{eqnarray}
  {\cal D}(\left\{x_{ij}\right\}) \equiv
  & {\cal D} &\left( x_{11},x_{12},\ldots,x_{1n}, \right.
  \nonumber \\
  && ~x_{21},x_{22},\ldots,x_{2n}, 
  \nonumber \\
  && ~\vdots
  \nonumber \\
  && \left. ~x_{N1},x_{N2},\ldots,x_{Nn} \right) \, .
\end{eqnarray}
The probability distribution ${\cal D}$ may not directly represent
data, but may refer to the posterior distribution obtained
from a previous inference, and the methods described below are
essentially unchanged.

It will be useful to classify models according to the dimensionality
of their outputs. (This dimensionality does not depend on the data,
except that it is specified as a probability distribution as described
above.) A ``zero-dimensional'' model, ${\cal M}$, with $M$ parameters,
$\{p\}\equiv p_1,p_2,\ldots,p_M$, is a model which produces a set of
$N$ predictions for each of the $N$ quantities of interest,
$\hat{x}_1=M_1(\{p\})$, $\hat{x}_2=M_2(\{p\})$, $\ldots$,
$\hat{x}_N=M_N(\{p\})$. In this case, the conditional probability
$P({\cal D}|{\cal M})$ is equal to ${\cal D}(\hat{x}_1,\hat{x}_2,
\ldots,\hat{x}_N)$. This is the case for which one can employ a
standard chi-squared analysis (in that case the quantities of interest
are the ``$y$'' values and the ``$x$'' values simply provide an index
for the data points which is irrelevant for the fit). The frequentist
best-fit is the maximum of ${\cal D}(\hat{x}_1,\hat{x}_2,
\ldots,\hat{x}_N)$ over the full model parameter space, and if ${\cal
  D}$ is of the form of Eq.~\ref{eq:mgauss} then the standard error
analysis applies.

A ``one-dimensional'' model produces a probability density function
for each point in parameter space rather than a unique value for each
quantity of interest, i.e. ${\cal M}_1(x_1,\{p\})$, ${\cal
  M}_2(x_2,\{p\})$, $\ldots$, ${\cal M}_N(x_N,\{p\})$. In this case,
the conditional probability is the $N$-dimensional integral
\begin{equation}
  P({\cal D}|{\cal M}) = \int \prod_{i=1}^{N}
  \left[ {\cal M}_i(x_i,\{p\}) d
    x_i\right] {\cal D}(x_1,x_2,\ldots,x_N) \, .
  \label{eq:m1d}
\end{equation}
As is typical in Bayesian inference, a prior distribution over the
model parameters must be specified as a separate probability
distribution, $P_{\mathrm{prior}}(\{p\})$, and then the posterior
distribution for, e.g., $p_1$ is
\begin{equation}
  P(\hat{p}_1) \propto \int \left( \prod_{i=1}^N dp_i \right) \delta
  \left(p_1 - \hat{p}_1 \right) P({\cal D}|{\cal M}) P_{\mathrm{prior}}(\{p\})
  \label{eq:marg} \, .
\end{equation}
An example of a one-dimensional model is a zero-dimensional model
which has a fixed Gaussian uncertainty, $\varepsilon$, in its
predictions, $\hat{x}_i$, for all of the quantities of interest.
Presuming the data points have only Gaussian correlations of the form
in Eq.~\ref{eq:mgauss}, then likelihood in Eq.~\ref{eq:m1d} can be
written in the form
\begin{eqnarray}
  P({\cal D}|{\cal M}) &\propto& \int \prod_{i=1}^{N}
  \left\{ \mathrm{exp}\left[-\left(x_i-\hat{x}_i\right)^2
    /\left(2\varepsilon_i^2\right) \right] d
  x_i\right\} \nonumber \\
  && \times
\mathrm{exp}
  \left[ - \frac{1}{2} (x_i-\mu_i) \Sigma_{ij}^{-1} (x_j-\mu_j)
    \right] \nonumber \\
  && = A \:
  \mathrm{exp}
  \left[ - \frac{1}{2} (\hat{x}_i-\mu_i) \hat{\Sigma}_{ij}^{-1}
    (x_j-\hat{\mu}_j) \right],
  \label{eq:ex1d}
\end{eqnarray}
where $\hat{\Sigma}_{ij} \equiv \Sigma_{ij}+\delta_{ij}
\varepsilon_i^2$ and $A$ is a normalization constant.
Since the final result is a single multivariate
Gaussian, Bayesian inference using one-dimensional models can be
represented by a Gaussian process. Eq.~\ref{eq:ex1d} derives the
result given in Eq.~2.20 in Ref.~\cite{Rasmussen06}. The frequentist
best-fit (i.e. the maximum of the likelihood function), in the case of
a one-dimensional model, is the maximum of the RHS of Eq.~\ref{eq:m1d}
over the parameter space, a maximization which requires, in general,
the evaluation an $N$-dimensional integral over each point. After
using the $\delta$ function to perform one of the integrations, the
computation of the posterior distribution in Eq.~\ref{eq:marg} is an
$N+M-1$ dimensional integral (though there are still only $M$ model
parameters). Joint posteriors for multiple model parameters can
be handled with multiple $\delta$ functions as above.

A two-dimensional model generates two-dimensional probability
distributions over pairs of quantities of interest. The conditional
probability is
\begin{eqnarray}
  P(D|M) \propto \int \prod_{i=1}^{N} \left[
    M_i(x_{i1},x_{i2},\{p\})~d x_{i1}~d x_{i2}\right] \nonumber
  \\ \times {\cal D}(x_{11},x_{12},x_{21},x_{22},\ldots,x_{N1},x_{N2})
  \, ,
  \label{eq:m2d}
\end{eqnarray}
which is a $2N$-dimensional integral. Note that the pair of variables
$(x_{11},x_{12})$ need not be the same as the pair $(x_{21},x_{22})$
and heterogeneous two-dimensional data sets may be fit using
Eq.~\ref{eq:m2d}. Generalizations to higher dimensions (or problems
with mixed dimensionality) are easily obtained.

The discussion above does not exhaust all of the potential
possibilities, as a model may generate a manifold embedded in some
higher dimensional space. The simplest example is a one-dimensional
model which generates a curve embedded in a two-dimensional data space
for each point in parameter space. The likelihood function, in this
case, is
\begin{eqnarray}
  P(D|M) & \propto \int_{c_i(\{p\})} & \left\{ \prod_{i=1}^N
  \left[d{\lambda}_i \left( g_{jk\,i} \frac{d x_{i}^j}{d{\lambda}_i} \frac{d
      x_{i}^k}{d{\lambda}_i} \right)^{1/2}
    \right] \right. \nonumber \\ && \times \left. M_i({\lambda}_i) {\cal
    D}[x^1_{i}(\lambda_i),x^2_{i}(\lambda_i)] \right\} \,
  \label{eq:curvefit}
\end{eqnarray}
where $c_i$ is the $i$-th curve (one curve for each data point and $N$
total data points) which the model produces given the parameter set
$\{p\}$, ${\lambda}_i$ specifies the line element along the $i$-th
curve, $g_{jk\,i}$ represents the $i$-th metric for the line integral,
$M_i({\lambda}_i)$ is the probability distribution along the $i$-th
curve which is specified by the model, and ${\cal
  D}[\left\{x_{i}(\lambda_i)\right\}]$ is the probability given by the
data, evaluated at the point $\lambda_i$ on curve $c_i$. Indicies $j$
and $k$ take values 1 and 2 and are summed over since they are
repeated and generalizations to larger than data spaces with
larger than two dimensions are straightforward. 

As might be expected from the discussion above, this is an
$N$-dimensional integral. In order to form posterior distributions
from this conditional probability one must also perform a sum over all
possible curves allowed by the model. In physical problems, one
typically parameterizes the curve with a finite number of parameters
and then posterior distributions are obtained by performing the
associated integral as in Eq.~\ref{eq:marg} above.

Note that, since the expression above employs $x(\lambda)$ and
$y(\lambda)$ but not $y(x)$ or $x(y)$, there is no requirement that
the curves are functions. The curves need not be continuous (the
functions $x(\lambda)$ and $y(\lambda)$ need not be continuous or
differentiable in order for the line integral to have meaning). This
means that model curves which ``go through'' the data multiple times
carry more weight, unless the model (or prior distribution) disfavors
such a scenario. Models which parameterize their predicted curves with
a finite number of parameters often disfavor arbitrarily complicated
curves. Also, models may vary the weight they assign to curves with
longer lengths by ensuring that the weight, $M_i(\lambda_i)$, is
proportional to $\ell_i^{-\alpha}$ given a finite curve length $\ell$
and some number $\alpha>1$.

The surprising aspect of curve fitting is the appearance of the
metric, $g_{jk}$, for each data point. In the simple case where
$g_{jk}=\delta_{jk}$, the integrand contains the usual line element
used to compute arc length. One intuititve way to see the metric
ambiguity, in the $N=1$ and $n=2$ case (a fit of one data point over a
two-dimensional space with a one-dimensional model), is to imagine
that $x_{11}$ and $x_{12}$ have different units. In that case, the
line integral appears nonsensical because the units cannot properly
cancel under the square root. The line integral is invariant under
reparameterizations of the curve but not under an arbitrary rescaling
of the coordinates. An alternative way to see that the metric required
is to note that the model curve may also be specified by a
$M=\delta\left[x_2-g(x_1,x_2)\right]$ and the transforming this form
to one similar to the one above requires a derivative $|\partial
g/\partial x_1|$ which contains the ambiguity represented by the
metric. Finally, the appearance of the metric is not solely
mathematical. For example, the process of fitting a theoretical curve
to experimental measurement of a particle's location at several points
(with some uncertainty in the coordinate directions), without any
other additional information, is possibly modified by black holes
which have traveled near the particle of interest and modified the
nature of space-time near the trajectory.

The line elements, denoted $\{\lambda\}$, are nuisance parameters which
must be integrated over. They also have prior distributions, which
must be specified to complete the inference. The problem of Bayesian
inference from this likelihood results in three possibilities: (i) the
model may specify what metrics ought to be used (leaving the
prior unspecified), (ii) the model may leave the metrics
unspecified and then then the choice of metric is also a prior choice,
and (iii) the metric may be unnecessary or trivial because of some
special feature of the model (e.g. if the quantities $x_{ij}$ have the
same units for all $j$). Options (i) and (ii) are related
because of the product of the likelihood and prior is unchanged when
multiplicative factors are moved between the two.
Without some additional simplification, our model now
effectively has $N+M$ parameters, the $M$ quantities $\{p\}$ and the
$N$ line elements $\{\lambda\}$.

One might suggest that one can obtain the likelihood for the curve
through a limiting procedure applied to the space in which it is
embedded, and thus avoid the ambiguity from the metric. However, the
limiting procedure is not unique. This is the Borel-Kolmogorov
paradox~\cite{Kolmogorov56,Jaynes03}, and the specification of
the metric in Eq.~\ref{eq:curvefit} avoids this paradox.

The conditional probability in Eq.~\ref{eq:curvefit}
reproduces a traditional chi-squared fit in
the case that the data has no uncertainty in one direction.
For example, if
\begin{eqnarray}
  &{\cal D}(x_{11},x_{12},x_{21},x_{22},\ldots,x_{N1},x_{N2})&
  \nonumber \\
  &\propto \prod_i \exp \left( -\frac{x_{1i}-\mu_i}{2
    \sigma_i^2}\right)
  \delta(x_{2i}-\eta_i)& \, ,
\end{eqnarray}
then one can choose $g_{jk\, i}=\delta_{jk}$ for all $i$,
${\lambda}_i=x_{2i}(x_{1i})$, and use the delta functions in ${\cal
  D}$ to do the line integrals. The factor $M({\lambda}_i)$ only serves to
reweight the data points and can be removed, finally leading to the
conditional probability for a typical chi-squared fit.

A frequentist equivalent could select the model which maximizes
$P({\cal D}|{\cal M})$ over the parameter space. In the case of curve
fitting (Eq.~\ref{eq:curvefit}) This optimization problem is itself
difficult, and the subject of algorithms designed to perform this
optimization~(e.g. Ref.~\cite{Fitzgibbon99}). In our Bayesian approach
where the integrals are often computable, in principle, using Monte
Carlo methods, but specific problems may result in functions which are
difficult to integrate.

A simpler alternative approach to Eq.~\ref{eq:curvefit} is to replace
each line integral in with a maximization. The alternative
likelihood of a curve labeled $c$ is
\begin{equation}
  {\cal L}_{\mathrm{alt}}(c) =
  \prod_i^N \mathrm{Max}_{{\lambda}_i}\left[{\cal
      D}_i(x_1,x_2,\ldots,x_n) M({\lambda}_i),c\right] \, ,
\end{equation}
where the $\mathrm{Max}$ function picks out the point along the curve
$c$ which has the largest value of the product ${\cal D} M({\lambda}_i)$.
However, the replacement of the line integral by a maximization forces
one to give up on coherence since some possible combinations of model
parameters and values for the parameters of interest are arbitrarily
removed from the posterior distribution~\cite{deFinetti90}. 

Eq.~\ref{eq:curvefit} presumes that the model predicts the shape of
the curve with zero uncertainty. This may not be the case, and the
uncertainty may increase the dimensionality of the problem leading to
fitting a surface (or manifold) to data rather than just a curve. When
the model generates a manifold, the formalism above can be easily
generalized, except that the metric on the manifold requires a prior
specification corresponding to the dimensionality of the manifold. For
example, a model which generates a two-dimensional surface gives a
surface integral over the data and requires a prior distribution with
two additional degrees of freedom corresponding to the two dimensions
in the surface.

\begin{figure}
  \includegraphics[width=3.1in]{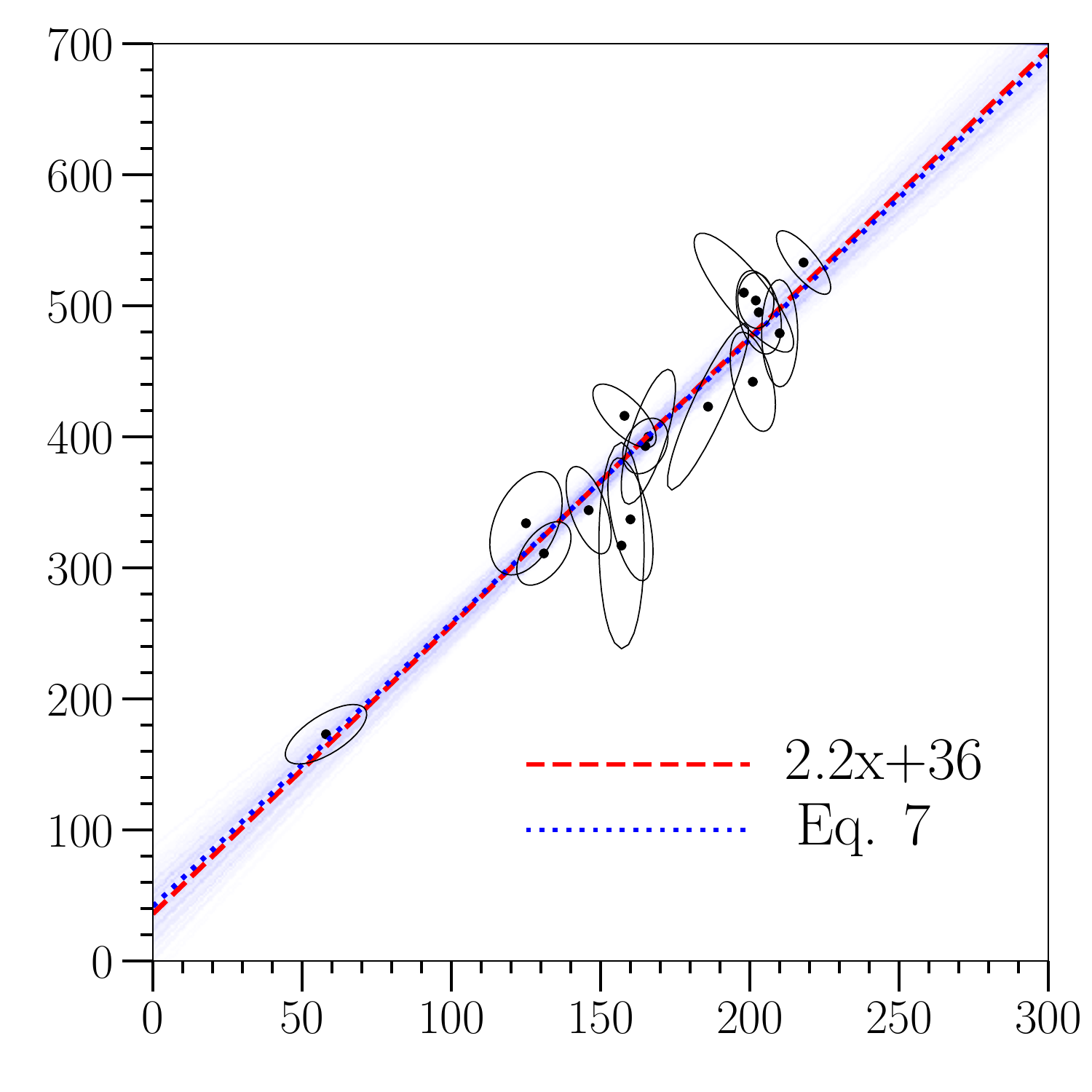}
  \caption{A fit of the test data from Ref.~\cite{Hogg10}, the best
    fit $y=2.2x+36$, and the best fit from an MCMC simulation of
    Eq.~\ref{eq:curvefit} (plotted with the same axis limits as
    Ref.~\cite{Hogg10}). For a linear fit, the metric term $(\sum
    dx^2/ds^2)$ just becomes a constant and can thus be ignored. As
    suggested in Ref.~\cite{Hogg10}, uniform priors were chosen in the
    y-intercept and angle, $\theta$, that the line makes with the
    positive x-axis. The underlying density plot shows samples from
    the posterior distribution from the MCMC simulation using
    Eq.~\ref{eq:curvefit}.}
\end{figure}

The uncertainty in the curve ${\lambda}_i$, parameterized by some quantity
$\eta_i$ may also lie in either the $x_{i1}$ or $x_{i2}$ directions.
In this case, the model must specify the probability distribution for
the set of possible curves and how to sum over this set. When the
model makes this specification, it is effectively specifying the
relationship betwen $({\lambda}_i,\eta_i)$ and $(x_{i1},x_{i2})$. Thus
any integration over ${\lambda}_i$ and $\eta_i$ can be transformed to
an integration over $x_{i1}$ and $x_{i1}$ and the conditional
probability reduces to the form given in Eq.~\ref{eq:m2d}. 

Similar curve fitting problems have been the subject of work in
frequentist data analysis, leading to several different approaches,
much of it focused on the case when the data points are nearly
normally distributed in both the $x_{i1}$ and $x_{i2}$ directions and
the curve to fit is nearly linear. In this case, one can use
traditional least squares in either the x- or y-direction, total least
squares, orthogonal least squares (a special case of total least
squares), or geometric mean (or reduced major axis)
regression~\cite{Stromberg40}. Traditional least-squares in the 
y-direction, for example,
corresponds to the choice $g_{jk}
= \bigl( \begin{smallmatrix} 0 & 0 \\ 0 & 1
\end{smallmatrix}\bigr)$, $\lambda=y$, and ignores the variation of the data
in the $x$ direction. Orthogonal least squares does not simply map on
to Eq.~\ref{eq:curvefit}, but corresponds to using a rotation matrix
for the metric with an orientation dictated by the slope of the curve
near each data point and then ignores the data variation orthogonal to
the rotation matrix. Ref.~\cite{Isobe90} compares related methods for
a particular class of problems from a frequentist perspective. None of
these methods, however, applies to the most general case when the
curves are highly nonlinear.

There are previous works which cover curve fitting in the Bayesian
perspective, most again on Gaussian distributed uncertainties where
the variance in the two coordinate directions naturally provides a
scales which help define the metric. Ref.~\cite{Florens74} describes
line fitting for errors-in-variables models and describes the
conditions under which the solution is not possible due to matrix
singularities. (The likelihood in Eq.~\ref{eq:curvefit} will similarly
exhibit difficulties if the metric is singular. Ref.~\cite{Frohner95}
performs a fit assuming a nearly linear curve, and uses the
approximation that the line integral over the probability distribution
from the data is approximately Gaussian. Ref.~\cite{dAgostini05} also
analyzes the nearly linear case with Gaussian uncertainties in the
data. Ref.~\cite{dAgostini05} considers the an additional uncertainty
in the $y$ direction and obtains a modification similar to the result
in Eq.~\ref{eq:ex1d}. None of these works describes the general result
in Eq.~\ref{eq:curvefit} above. Ref.~\cite{Pihajoki17ag} describes a
complementary geometric method to fitting lower-dimensional models to
data in the context of Bayesian inference. The discussion above,
however, is unique in its identification of the ambiguity from the
metric and the observation that the metric choice may need to
be part of the prior.

\begin{figure}
  \includegraphics[width=3.1in]{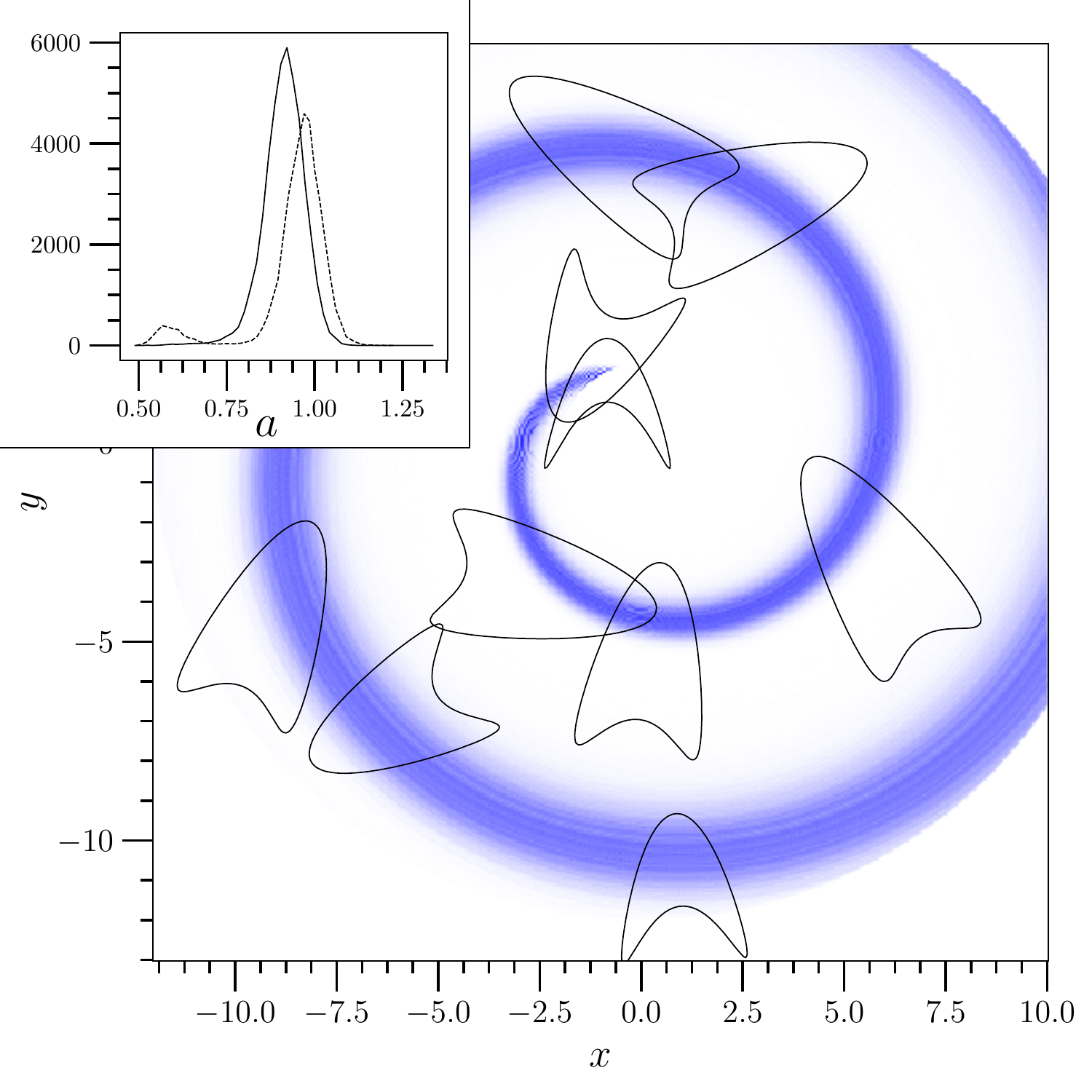}
  \caption{A fit of sample data to a one parameter model $r=a \theta$
    which could not be easily performed using traditional regression
    techniques. The parabolic contours give 68\% contours for the 10
    data points, the density plot shows the posterior of the fit with
    trivial metric for $\theta>1.6$, and the inset compares the
    posterior distribution for the parameter $a$ using either the
    trivial metric (solid line) or modified metric (dashed line). The
    prior probability distributions for $a$ and $\lambda=\theta$ are
    taken to be uniform for both metrics.}
\end{figure}

Ref.~\cite{Hogg10} describes the Bayesian generalization of orthogonal
least squares, and fits a sample Gaussian data set to a line to
demonstrate the approach (see figure 9 in Ref.~\cite{Hogg10}). In figure
1, the same data set is shown including an analysis with the likelihood in
Eq.~\ref{eq:curvefit}. A nearly indistiguishable result is obtained,
demonstrating that our formalism works with a simple problem. A more
physical example is in Ref.~\cite{Beloin18cs} where theoretical
neutron star cooling curves are compared with data. In this case, the
use of Eq.~\ref{eq:curvefit} is required because of large
uncertainties in the age observations. There is no clear model
guidance on the metric so a simple choice is made.

A more difficult fitting problem which demonstrates the method and
cannot be easily solved from the method in Ref.~\cite{Hogg10} is that
shown in figure 2 which is fit to a model curve $r=a \theta$ for one
parameter $a$ with a uniform prior distribution with $0<a<20$. The
data was generated from $r=a \theta$ with $a=1$ and independent noise
added in the x and y directions. The data are generated from the
function $f(x,y)=\exp[-(y-x^2)^2] \exp(-x^2/2)$ with translations to
nearly match $r=\theta$ and random rotations. The metric
$g_{jk}=\delta_{jk}$ is used and the curve is parameterized with
$\lambda=\theta$ (the line integral is invariant under
reparameterizations of $\lambda$). The choice of metric is a prior
choice and a uniform prior distribution in $\lambda$ is assumed. The
posterior model curves are shown as a density plot in figure 2 and a
posterior histogram for the parameter $a$ is given as the solid line
in the inset. Modifying the choice of metric to $g_{jk} =
\bigl( \begin{smallmatrix} 10^6 & 0 \\ 0 & 1
\end{smallmatrix}\bigr)$
gives a different posterior for $a$, shown in the dashed line
in the inset in figure 2. Only a large change in the metric is able to
overcome the exponential suppression in the data, but the histogram is
qualitatively modified. This is typical for curve fitting: the
posteriors for the model parameters depends (albeit weakly if the data
is sufficiently accurate) on the choice of the metric.

In summary, a new Bayesian approach to curve fitting has been
presented which applies to a wide variety of problems and suffers from
no ambiguity beyond the proper specification of the prior probability.
It gives the correct result in the limiting case of a straight line
fit. Our approach generalizes to higher dimensional cases where the
model produces a manifold embedded in a higher-dimensional space
occupied by the data. From the Bayesian perspective, much of the
frequentist literature on this problem attempting to optimize the
choice of frequentist method for the problem at hand is equivalent to
trying to optimize the choice of prior distribution. Thus, one expects
that objective Bayesian priors such as the Jeffrey's prior could be
applied to this problem, and this possibility is left to future work.

\section*{Acknowledgements} 

The author would like to thank George Bertsch, Ed Brown, Cole Miller,
and Pauli Pihajoki for useful discussions. This work was supported by
NSF grant PHY 1554876 and by the U.S. DOE Office of Nuclear Physics.
This work was initiated at the Aspen Center for Physics, supported by
NSF grant PHY 1607611.

\bibliographystyle{apsrev}
\bibliography{twod}

\end{document}